# Achieving High Yield of Perpendicular SOT-MTJ Manufactured on 300 mm Wafers

Wenlong Yang*, Zhenghui Ji*, Yang Gao, Kaiyuan Zhou, Qijun Guo, Dinggui Zeng, Shasha Wang, Ming Wang, Lijie Shen, Guilin Chen, Yihui Sun, Enlong Liu, Shikun He

Zhejiang Hikstor Technology Co. LTD., Chongwen Road 1718, 311305, Zhejiang, China.

Email: liuenlong@hikstor.com; he_shikun@hikstor.com

* These authors contributed equally to this paper.

*Abstract*—The large-scale fabrication of three-terminal magnetic tunnel junctions (MTJs) with high yield is becoming increasingly crucial, especially with the growing interest in spin-orbit torque (SOT) magnetic random access memory (MRAM) as the next generation of MRAM technology. To achieve high yield and consistent device performance in MTJs with perpendicular magnetic anisotropy, an integration flow has been developed that incorporates special MTJ etching technique and other CMOS-compatible processes on a 300 mm wafer manufacturing platform. Systematic studies have been conducted on device performance and statistical uniformity, encompassing magnetic properties, electrical switching behavior, and reliability. Achievements include a switching current of 680 µA at 2 ns, a TMR as high as 119%, ultra-high endurance (over $10^{12}$ cycles), and excellent uniformity in the fabricated SOT-MTJ devices, with a yield of up to 99.6%. The proposed integration process, featuring high yield, is anticipated to streamline the mass production of SOT-MRAM.

## I. Introduction

Spin-transfer-torque magnetic random access memory (STT-MRAM) has garnered significant attention as a promising memory technology due to its high speed, low power consumption, high endurance, and compatibility with CMOS processes [1, 2]. Despite making its way into themarket, the potential applications of STT-MRAM are limited by the incubation time and the shared read/write path [3, 4]. To address these limitations, spin-orbit-torque MRAM (SOT-MRAM) has emerged as a recent solution, featuring three-terminal magnetic tunnel junctions (MTJs) with separate read and write paths that offer sub-ns switching speed, unlimited endurance, and enhanced read stability[5-9]. This positions SOT-MRAM as a promising candidate for integration into last-level cache or other applications requiring high speed and reliability.

Typically employing a top-pinned structure, MTJs for SOT-MRAM involves the positioning of the free layer (FL) above the SOT channel, where the flow of charge current through the SOT channel generates spin current via the spin Hall effect or Rashba effect to facilitate FL switching [10-12]. So far, the development of SOT-MRAM is still in the research phase due to challenges in multiple manufacture processes, especially SOT-MTJ etching [13, 14]. Since the thickness of SOT channel is typically around 5 nm, it is crucial to precisely control the stopping position exactly on the SOT channel across the whole wafer. Meanwhile, the redeposited metal residues surrounding MTJ pillar will lead to electric short in single device without extra treatment and decrease device yield. Several previous works have focused on solving this issue. For example, a bi-layer bottom electrode (BE) structure was adopted and a 96% yield of resistance-based MTJ was obtained on 300 mm wafer due to the high selectivity between the MTJ and

SOT channel during MTJ etching [15]. However, there is no description of the device uniformity on wafer-level. Zhao *et al.* developed a stop-on-MgO etching technique, enabling a 100% yield of resistance-based MTJ by effectively preventing sidewall re-deposition around the MgO barrier [16]. However, achieving precise control of the stopping position on a 300 mm wafer during the ion beam etching (IBE) process remains challenging, considering the absence of selectivity between MgO and other magnetic materials. Developing an effective process integration solution to enhance SOT-MTJ yield on a 300 mm wafer platform is thus imperative.

In this paper, we successfully integrated SOT-MTJ devices with perpendicular magnetic anisotropy (PMA) on a 300 mm wafer platform using CMOS-compatible processes, and systematically studied the magnetic properties, electrical characteristics, uniformity and reliability performance. Through the developed integration process, a 99.6% yield was attained among fabricated SOT-MTJ devices on the wafer, showcasing a promising outlook for high-yield SOT-MRAM development.

## II. INTEGRATION OF SOT DEVICE

The integration of SOT-MTJ devices takes place at a 300 mm wafer manufacturing facility. The main steps of the integration process, as well as the schematic of a single MTJ device, are depicted in Fig. 1(a) and (b) respectively. The substrate comprises a bottom metal (BM), bottom via (BV) and bottom electrode (BE) to establish electrical contact with the SOT channel. The MTJ stack is deposited on smoothed BEs with a spacing of 200 nm. A specific MTJ etching method has been developed to precisely stop on the SOT layer without leaving any redeposited metal residues around the MgO barrier, which will be detailed below. The critical dimension of an MTJ pillar is approximately 80 nm. Subsequently, the SOT layer and top electrode (TE), each with a width of 200 nm, are simultaneously etched. This is followed by the fabrication of a dual damascene Cu contact (top via and top metal, TV and TM) and aluminum (Al) pads to electrically connect the top and bottom metal layers for single device testing from a 51×51 MTJ bit array. The MTJ integration process is accompanied by corresponding oxide filling and chemical-mechanical planarization (CMP) modules.

As illustrated in Fig. 1(c), the MTJ stack is top-pinned (TP) with PMA and consists of SOT layer/CoFeB/MgO/CoFeB/ synthetic-antiferromagnet (SAF). The film stack undergoes annealing at 350 °C for 30 minutes in vacuum after deposition. Based on current-in-plane tunneling (CIPT) measurement, the resistance-area product (RA) and tunnel magnetoresistance (TMR) are 36 $\Omega \cdot \mu m^2$ and 130% respectively. Tungsten (W) is utilized as the SOT layer due to its compatibility with CMOS processes and high SOT efficiency [17, 18]. By optimizing the growth conditions of W, the optimum properties are achieved at $t_W$ = 4 nm, with $\rho_W$ = 250 $\mu\Omega \cdot cm$ and $\theta_{SH}$ = -0.3. A typical cross-section transmission electron microscopy (TEM) image of an SOT-MTJ is displayed in Fig. 1(d).

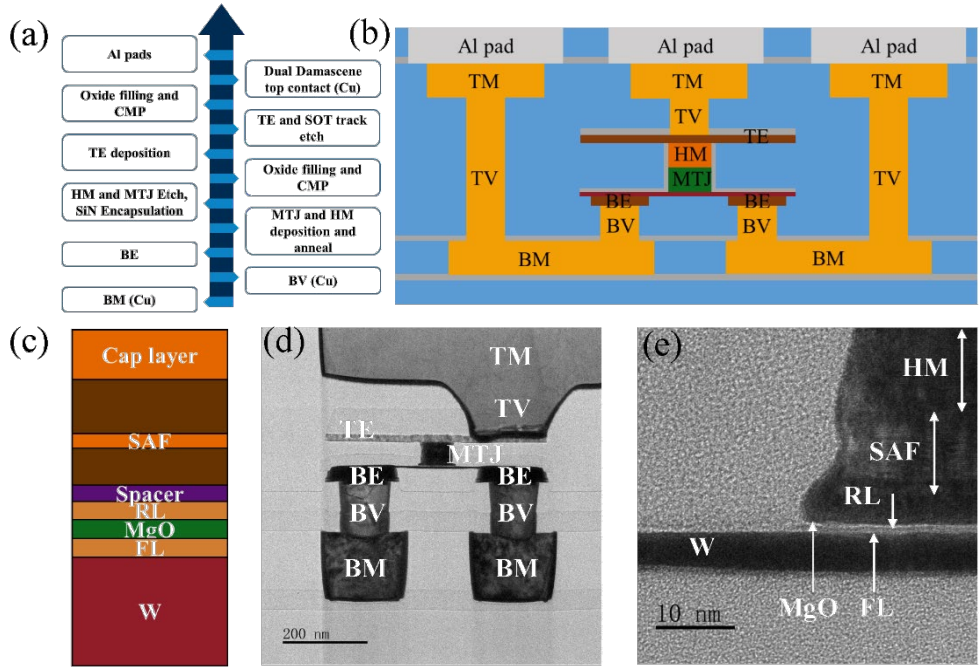

Fig. 1. (a) Main steps of integration process. (b) Schematic of a single SOT-MTJ device. (c) Schematic of TP film stack. (d) Cross-sectional TEM image of a three-terminal cell. (e) The high resolution TEM image on MTJ edge.

To address the etching issue of SOT-MTJ, a novel etching method is proposed as illustrated in Fig. 2(a)-(d). Firstly, the SOT-MTJ is patterned into a pillar through IBE and precisely stopped on the SOT track by monitoring optical emission spectroscopy (OES). Secondly, a SiN encapsulation layer is deposited to protect the MTJ from oxidation. Thirdly, a combination of etching at two different grazing angles ($\varphi_1$ and $\varphi_2$) is conducted to etch SiN due to the different step coverage between the bottom and sidewall regions. The sidewall etching rate was faster at angle $\varphi_2$ compared to angle $\varphi_1$, thereby ensuring the complete removal of SiN and metal residue from the sidewall, while a small amount of SiN in the bottom region remains. The insulating characteristics of SiN in the bottom region facilitates suppression of the metal sputtering phenomenon and thereby enhances the yield. Finally, a SiN encapsulation layer is deposited again to prevent oxidation of the MTJ. The high-resolution TEM image on the MTJ edge in Fig. 1(e) reveals the absence of re-deposited metal residues across the MgO barrier, indicating precise etching termination on the SOT track.

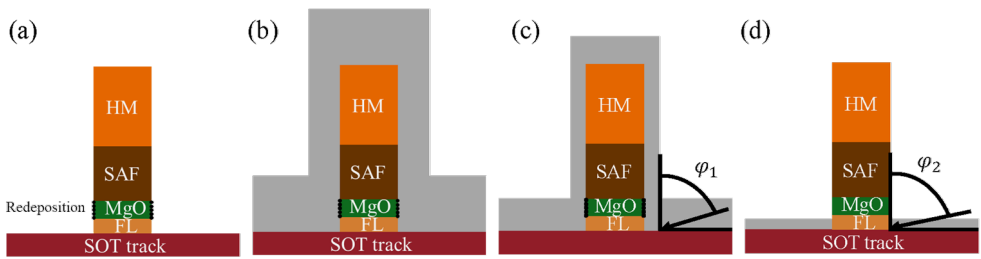

Fig. 2. Schematic of the IBE process for MTJ fabrication. (a) The MTJ is etched and stopped on the SOT track. (b) SiN encapsulation. (c) Etching at angle $\varphi_1$. (d) Etching at angle $\varphi_2$.

## III. RESULTS AND DISCUSSION

Fig. 3(a) illustrates a typical resistance-field (*R-H*) hysteresis loop from an SOT-MTJ device, showcasing a high TMR ratio of 119%. However, the TMR value is reduced by 11% compared to that of the film stack due to etch-induced damage. To characterize the distribution of SOT-MTJ properties, the same loop and SOT track resistance tests were performed on a total of 729 devices across the entire wafer. The average values for FL coercive field ($\mu_0 H_c$), offset field of FL ($\mu_0 H_{offset}$), TMR, MTJ resistance at parallel state ($R_P$), and $R_{SOT}$ are measured at 75 mT, 0.36 mT, 119%, 10.89 kΩ, and 776 Ω respectively, as depicted in Fig. 4(b)-(e). The variation coefficient (CV) of $\mu_0 H_c$, TMR, $R_P$, and $R_{SOT}$ is found to be 8.98%, 3.21%, 13.27%, and 7.19%, respectively. These basic magnetic characterizations at the wafer level suggest a relatively uniform performance in the 300 mm wafer manufacturing platform, although further optimizations are deemed necessary for specific process in the proposed integration to enhance magnetic properties and uniformity.

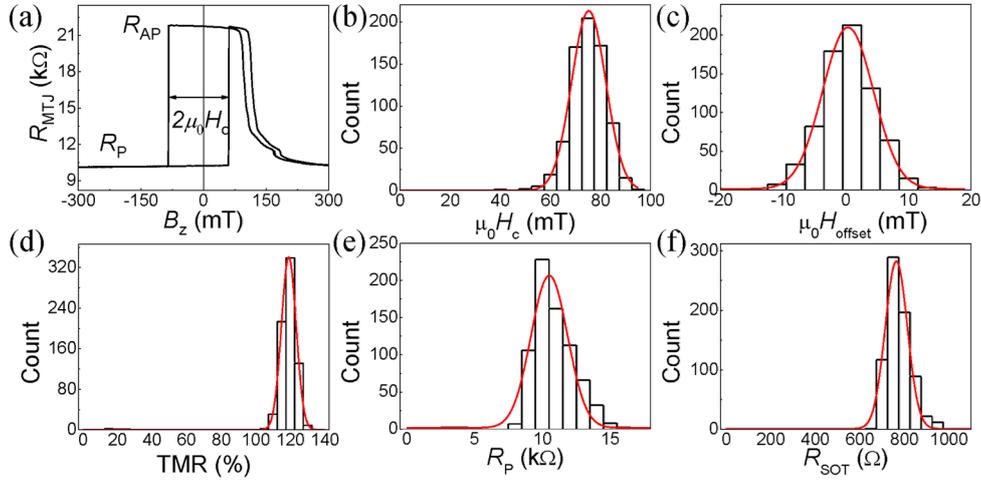

Fig. 3. (a) A typical *R-H* hysteresis loop of SOT-MTJ device. The statistical data are shown for (b) $\mu_0 H_c$, (c) $\mu_0 H_{offset}$, (d) TMR, (e) $R_p$ and (f) $R_{SOT}$ extracted from devices throughout the full wafer.

During electrical switching measurements, an external magnetic field ($B_{ext} = \pm 20$ mT) is applied along the current flow direction. As revealed in Figure 4(a), the chirality of SOT switching changes with the direction of the external magnetic field, which is consistent with SOT switching characteristics. The average critical switching current of 234 devices as a function of applied pulse widths ranging from 100 ns to 2 ns is shown in Fig. 4(b). It is noteworthy that the critical switching current gradually decreases with increasing pulse width. Additionally, differences in the critical switching current of $I_{P-AP}$ and $I_{AP-P}$ are observed due to stray field in the vertical direction of SAF and electromagnet, where $I_{P-AP}$ ($I_{AP-P}$) represents the critical switching current from P (AP) to AP (P) state. Due to the inherent limitations of our pulse generator, we are unable to demonstrate the switching behavior exhibited by shorter pulses. However, it is worth emphasizing that the average critical switching current values at 2 ns are noted as 680 μA for $I_{P-AP}$ and 880 μA for $I_{AP-P}$, corresponding to current densities of 0.85 ×$10^{12}$ A/m$^2$ and 1.1 ×$10^{12}$ A/m$^2$, respectively, which are smaller compared to reported values [5, 9]. Further reduction in critical switching current, from a CMOS scaling perspective, can be achieved by scaling the width of the SOT track, since there is still plenty of room for track width to bring down (200 nm) compared to the CD of MTJ (80 nm). Moreover, Fig. 4(c) summarizes the critical switching current distribution of SOT-MTJ devices at 5 ns across the entire wafer, displaying a considerable range that poses challenges for integration with CMOS technology. Further process integration optimization is necessary to address this issue.

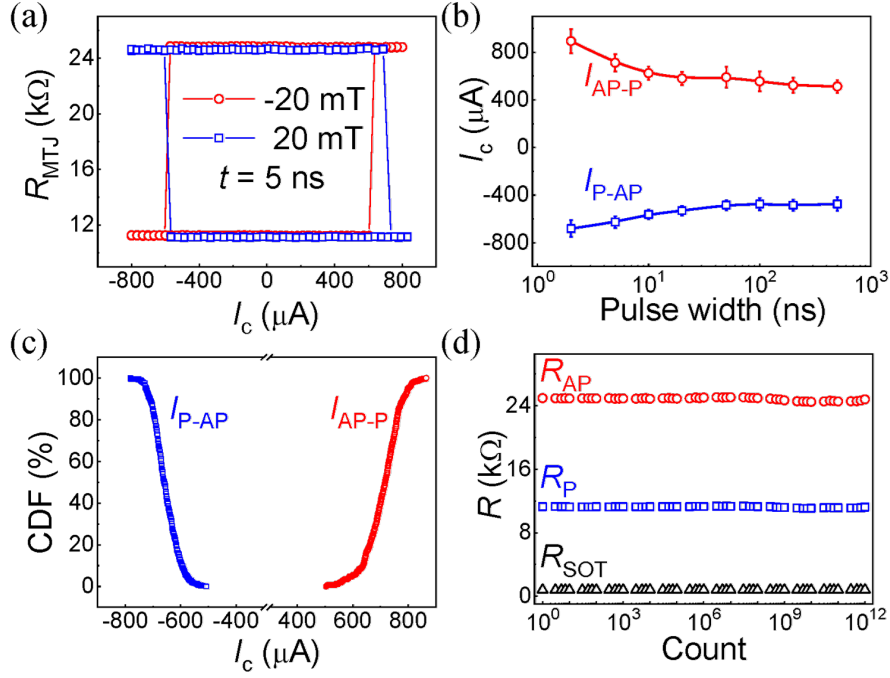

Fig. 4. (a) SOT induced magnetization switching with an in-plane magnetic field ($B_{ext} = \pm 20$ mT). (b) The average critical switching current for various pulse widths. (c) Cumulative distribution function (CDF) plot of the critical switching current at 5 ns. (d) The measured values of $R_{AP}$, $R_P$, and $R_{SOT}$ after cycling with the corresponding voltage pulses.

To assess reliability performance, endurance tests by applying voltage pulse with a duration of 5 ns and an amplitude of 0.8 V were conducted on single SOT-MTJ devices randomly chosen from fabricated wafers. Resistance changes in $R_{AP}$, $R_P$, and $R_{SOT}$ monitored to evaluate device performance are depicted in Fig. 4(d) from one of the tested SOT-MTJ devices. The endurance test, limited to $10^{12}$ cycles due to time consumption, showed no significant resistance changes after write cycles, underscoring the advantages of utilizing the proposed integration process method for the fabrication of SOT-MRAM devices.

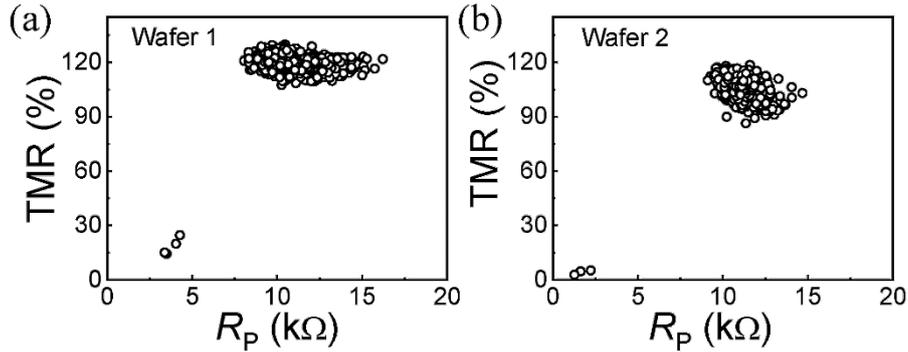

Fig. 5 Scatter plots of TMR-$R_P$ for (a) wafer1 and (b) wafer 2 devices.

To further investigate yield, TMR-$R_P$ scatter plots are depicted for two wafers in Fig. 5(a) and (b). Although the preparation process for these wafers varied slightly during the etching step, both wafers encompass 729 devices in total. Wafer 1 demonstrated 4 short bits, while wafer 2 displayed 3 short devices, thus achieving a 99.6% yield across all devices on the entire wafer. The results from these two wafers highlight the reproducibility of the preparation process. While obtaining more precise yield metrics necessitates detailed statistical analysis of a device array, the above results signify the attainment of low damage and high yield through the proposed integration process method, showing significant potential for SOT-MRAM chip fabrication on 300 mm wafers.

## IV. CONCLUSION

In summary, we propose a novel fabrication method for achieving high-yield (up to 99.6%) preparation of perpendicular magnetized SOT-MRAM devices, which has been successfully implemented on a 300 mm process platform. Furthermore, we systematically investigate the magnetic properties, electrical characteristics, uniformity and reliability performance of these devices within the 300 mm wafer manufacturing platform, including low critical switching current (680/880 μA at 2 ns), high TMR (119%), ultra-high endurance (over $10^{12}$). Most importantly, the $R_{SOT}$, $R_P$, and TMR ratios exhibited a relatively good uniformity with variation coefficients of 7.19%, 13.27% and 3.21%, respectively. These results indicate that the proposed process integration solution exhibits significant potential for advancing high-yield SOT-MRAM development.

## ACKNOWLEDGMENT

This work is supported by National Science and Technology Major Project (2020AAA0109003). We also acknowledge the support from Hikstor's pilot line.